\documentclass[runningheads]{llncs}
\usepackage{graphicx}
\graphicspath{ {./images/} }
\usepackage{amsmath} 
\usepackage{booktabs}
\usepackage{float,lscape}
\usepackage[utf8]{inputenc}
\usepackage{multirow}
\usepackage{csquotes}
\usepackage[english]{babel}
\usepackage{hyperref}
\usepackage{url}
\usepackage{array}
\usepackage{float}

\usepackage[backend=biber]{biblatex}
\begin{document}
\title{Norm violation in online communities \textendash  A study of Stack Overflow comments}

\author{Jithin Cheriyan\and
Bastin Tony Roy Savarimuthu\and
Stephen Cranefield}
\authorrunning{Jithin Cheriyan et al.}
\institute{Department of Information Science, University of Otago, Dunedin, New Zealand \email{\{jithin.cheriyan,tony.savarimuthu,stephen.cranefield\}@otago.ac.nz}}

\maketitle            

\begin{abstract}
Norms are behavioral expectations in communities. Online communities are also expected to abide by the established practices that are expressed in the code of conduct of a system. Even though community authorities continuously prompt their users to follow the regulations, it is observed that hate speech and abusive language usage are on the rise. In this paper, we quantify and analyze the patterns of violations of normative behaviour among the users of Stack Overflow (SO) \textendash  a well-known technical question-answer site for professionals and enthusiast programmers, while posting a comment. Even though the site has been dedicated to technical problem solving and debugging, hate speech as well as posting offensive comments make the community ``toxic". By identifying and minimising various patterns of norm violations in different SO communities, the community would become less toxic and thereby the community can engage more effectively in its goal of knowledge sharing. Moreover, through automatic detection of such comments, the authors can be warned by the moderators, so that it is less likely to be repeated, thereby the reputation of the site and community can be improved. Based on the comments extracted from two different data sources on SO, this work first presents a taxonomy of norms that are violated. Second, it demonstrates the sanctions for certain norm violations. Third, it proposes a recommendation system that can be used to warn users that they are about to violate a norm. This can help achieve norm adherence in online communities.

\end{abstract}

\section{Introduction}
Online social media platforms have enabled users to express their viewpoints and hence have become a place for information sharing. Applications like Facebook and Twitter are the  forerunners in this arena along with multitudes of other applications. Interactions among users of these applications are generally observed as positive, inclusive and creative. SO, a technological division of Stack Exchange --- a network of question-answer websites on topics in diverse fields, is a platform for beginners to get free technical support from professionals and well-versed programmers. Those who are interested can join the site for free and post questions they may have about programming to get the best possible theoretical and practical support. Anyone can ask questions in any joined communities and anyone can post answers or comments to that question, making the site dynamic and inclusive. The intent of SO is to give power back to the community~\cite{noauthor_who_nodate} so that a new way of knowledge sharing can be created.

This work is inspired by the post by the Executive Vice President of SO, regarding the alarming transformation of SO as an unwelcoming place~\cite{culture_stack_2018}. Even though millions of comments are generated by the users of various communities day-by-day, a considerable amount of them were found to violate the Code of Conduct (CoC) of the site~\cite{noauthor_code_nodate}, which advocates for friendliness and inclusiveness.  To monitor the proper usage of the site, site authorities have selected reputed community members as moderators to monitor and review all the posts~\cite{culture_stack_2018}. The reputation score in SO decides a user's future as a moderator. Reputation score comes from a range of activities including the up-votes of all the answers that one makes and it reflects how much a person has been accepted as a valued resource in that community~\cite{noauthor_what_nodate}. If a comment is found to breach the CoC of the site, moderators would either remove that comments or may contact the author to remove it. Some examples of comments that have been deleted by the moderator are given below.

\begin{flushleft}

{\itshape``shut up sir....."}\\
{\itshape``I just hate this answer oh downvote you senseless clods \textless  profanity \textgreater."}\\
{\itshape``Its called your brain. If you can't review your code ask someone else to do it."}\\
{\itshape``you can convert it into seconds, then compare. I think you learned that in your school."}

\end{flushleft}

In addition to these human moderators,  SO brings into play an automatic bot~\cite{noauthor_app_nodate} which helps to identify comments containing certain triggering keywords reflecting toxic contents, and will report that to the moderators if they exceed a certain threshold. These norm violating comments, bearing highly toxic contents, would be flagged `red' and those will be moderated by the human moderators immediately~\cite{co-founder_raising_2009}. Moderators who are online will be sent a notification to deal with this. Thereby an important set of norms are enforced on the site. Figure 1 shows the process of moderation in SO. However, no prior work has investigated the types of norms and their violations pertaining to SO and this work bridges this gap. 

\begin{figure}
\includegraphics[width=\textwidth]{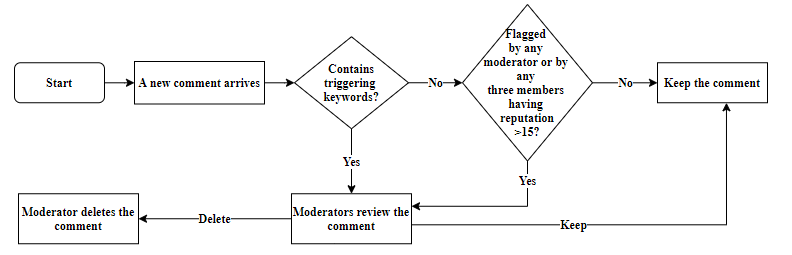}
\caption{ A flowchart of comment moderation in Stack Overflow} 
\end{figure}

Even though SO has been offered in myriads of languages like Spanish, Portuguese, Russian and Japanese~\cite{noauthor_posts_nodate6}, this work intends to study the norm violations in comment posting in English only~\cite{co-founder_non-english_2009}. In this proposed work, by analysing comments on SO, we address three objectives: 1) propose a classification of the types of norms and also quantify norm violations, 2) quantify the different types of punishments for norm violations and 3) propose a recommendation system to minimise norm violations in SO in terms of comment posting.

The paper is structured as follows. Section 2 provides background and related work in norms, norm identification and norm violations. Also, this section establishes the purpose of identification of norm violation in online discussion forums. Section 3 describes the context of norm investigation process in this study. The methodology is explained in Section 4 and Section 5 presents our results. Finally, Section 6 presents the implications of the study in the context of a normative recommendation system and Section 7 concludes the paper.                                                                                                                                             

\section{Background and related work}
In many areas of human life,  rules, conventions and norms play a vital role in the smooth functioning of the system~\cite{Savarimuthu1,sen2007emergence}. Even though rules are enforced by law or authorities, norms are considered as the set of expected practices of interpersonal behaviour in societies, and may only be socially monitored~\cite{Savarimuthu1}. While the breaching of rules results in punishment, usually norm violations may not get punished all the time~\cite{Savarimuthu1,Savarimuthu2}. As norms are not been imposed by those in power, it is the mindset of the people and the extent to which sanctions are perceived to be applied which make members of society follow or violate the norms. Since norms are specific to a society, identifying and following a certain norm in an unknown community is challenging for people~\cite{Gao,Savarimuthu1,savarimuthu2017developers,sen2007emergence,singh2013uses}. A system that is able to recognise potential norm violations and warn users about them would be ideal. However, such a system must be able to identify what are the norms that exist in the society.

Many researchers have investigated the norm creation, norm learning and norm emergence processes in agent systems, especially in multi-agent systems (MAS)~\cite{Savarimuthu1,savarimuthu2013social,sen2007emergence}. Other than research on agent based systems, many others have focused on norm adherence and violation by the members of various social media platforms like Facebook and Reddit~\cite{chandrasekharan2019hybrid}. For example, Chandrasekharan et al. have tried to identify and classify the macro and meso norms being violated in Reddit comments~\cite{Chandrasekharan2}. Nowadays, to identify and moderate online hate speech, for an example, a meaningful amalgamation of both these streams has been utilized by all leading social networking sites~\cite{park_one-step_2017,razavi2010offensive}. We examine both of these streams of norms identification in the following sub-sections.

\subsection{Norms in multi-agent systems}
MAS may contain both artificial and human agents. Therefore, it is expected that these communities also would follow certain behavioural norms inside the community. Usually, agents follow the norm life cycle --- norm creation, spreading, learning, enforcement and emergence~\cite{Gao,Savarimuthu1,Savarimuthu4}. Norms are created by the norm-leaders and inferred by agents by observing the patterns of actions of other agents of the society.  Thus, that predominant practice would become a norm  to be followed by the community. 

In a MAS, the learning process can be either offline or online~\cite{Savarimuthu1,singh2013uses}. The offline mode of learning, certain rules would be embodied into the agents and they would follow these regulations in a top-down model. But, in a MAS, the norms may be changed dynamically, requiring an ability of an agent to learn the updated norm from the behaviour of other agents. Therefore, in the online fashion of learning, multiple agents interact with others simultaneously and at the same time they would learn the etiquette through the interactions. This is a bottom-up approach and it is usually expected that the agents should learn from their own experiences to fit in a dynamic community. Like human beings, agents also may communicate to achieve coordination and cooperation~\cite{Xuan}. Usually, as part of online learning, the agents may exchange information regarding their present state or the resources they hold, thereby gaining a better understanding of themselves and others' expectations of its behaviour~\cite{anastassacos_understanding_2019}.

Norm enforcement refers to the process of discouraging norm violation either by sanctions or punishments~\cite{Savarimuthu1}. Punishment could be monetary or blacklisting an agent. As per Singh et al.~\cite{singh2013uses} and Savarimuthu et al.~\cite{Savarimuthu1}, categories of norms are a) obligation norms  2) prohibition norms and 3) permission norms. Obligation norms are the set of patterns of actions which an agent is supposed to do like tipping in a restaurant. Likewise, an agent system is not expected to perform an action that has been prohibited by the society such as littering the park. Violating the above mentioned two classes of norms would invite sanctions or punishments while the third one, permission norms, usually may not get sanctioned as it refers to the set of actions an agent is permitted to do. This work concerns prohibition norms.

\subsection{Mining norms from online social media platforms}
Online social networking sites like Facebook, Twitter and Reddit provide a digital world where interactions happen. Even though there are centralised rules regarding the effective and constructive usage of these applications, there exist some users who do not abide by these rules, resulting in abuse and hate speech. When people get upset, especially in online media, they are likely to overlook the norms and may start bullying others for even small mistakes, which has found a surge in last decade~\cite{noauthor_alw_nodate}. Also, it is observed that aggressive behaviour is more common in online spaces than face-to-face~\cite{jones_online_2013}. This hostile nature has long lasting detrimental effects, e.g. the CEO of Twitter has admitted that Twitter loses their users because of online hate speech~\cite{noauthor_alw_nodate}. The European Union has passed a law that all leading social media platforms must delete abusive contents within 24 hours ~\cite{noauthor_alw_nodate}. This justifies the requirement of a scalable computational system to systematically locate and remove online hate speech in social media platforms.

As a result of social and legislative pressure, all prominent social media applications are forced to employ human moderators from within the community to identify and moderate provoking, abusive and unnecessary contents~\cite{corr2018legal}. These community moderators monitor the posts that have violated the norms of the community, and potentially delete these posts. Along with these human moderators, computational tools like chatbots (agents) are used to identify abusive or offensive terms and notify the human moderators for possible moderation. Abusive language detection is an interdisciplinary domain which integrates technical components like natural language processing and machine learning~\cite{gupta_natural_2019}. Even though the human moderation process has been found effective, this tedious task could be minimised if all the community members follow the norms. Unfortunately, norm violation in social media is becoming commonplace and there is an increased need for understanding the types of norms and their violations in online communications. In this work we investigate the norms of SO on a posted comment.

The goal of the paper is to identify the norms that a comment is expected to follow and the nature of violations. We provide the context of our study in the next section.

\section{Norm investigation in Stack Overflow}
Having learned about the toxicity through the aforementioned blog post of SO top officials, we investigated the nature of comments by collecting the comments from one particular day with the goal of checking the number of comments that are deleted in subsequent days. We found that around 9.3\% of total comments of that day disappeared within one month and this increased to 14\% within two months, and this process still continues. Comments were collected between 2 December 2019 and 13 February 2020 for posts with tags `Java' and `Python'. Between these two dates, 3221 comments that originally appeared on 2 December were deleted. There are two possible reasons for a comment's disappearance: either voluntarily deleted by the author or removed by the moderator as part of moderation (i.e. deleted because it violates a norm). Therefore, the set of deleted comments comprise those that were voluntarily removed by the author and deleted by the moderator. The reason for moderation could be the possible violation of generic rules of a society or, some rules or norms specific to SO. Based on prior work~\cite{al-hassan_detection_2019} and a bottom-up analysis of deleted comments, in this work we investigate two major classifications of norms and patterns of norm violations that can be observed from textual comments: (1) Generic norms and (2) SO specific norms. 

\paragraph{\bfseries Generic norms and their violations - } Generally, it is expected that in the community we live in, we are not supposed to use rude or offensive words that may hurt others. So, the norm is to be polite, and also respect individuals and their unique attributes such as gender and ethnicity. While interacting in online communities we are also expected to follow these rules to keep the propriety.  SO is such a forum where the CoC requires its users to respect fellow community members irrespective of their knowledge level, ethnicity or gender. Therefore, generic norms refer to a set of universally accepted norms pertaining to the use of refined language. These norms are likely to be similar across online communities. The norms under this category of prohibition norms are norm against a) personal harassment, b) use of racial slurs, c) use of swear words, and d) use of unwelcoming language. A fine grained description along with examples of these norms are given in Table 1. 
\paragraph{\bfseries SO specific norms and their violations - }It is a common courtesy to express gratitude for any kind of help, especially in online media. People will acknowledge and apologize for small mistakes to keep the space amiable. However, as aforementioned, the CoC of SO has the policy of keeping the site useful to everyone by containing only relevant knowledge and reducing noise. As a part of that, even though users usually post thanksgiving or apologising comments, these are usually removed by moderator as those comments may distract the users from gaining knowledge, or the newly joining members may consider the site to be too chatty~\cite{rahman_cleaning_2019}. Also, if someone asks an irrelevant question as a follow up of a posted question, this may also get removed. These norms of not accepting or not promoting pleasantries are unique to SO as these comments distract from the intended purpose of knowledge sharing. Therefore, the very specific norms of SO is to keep the site less noisy (free of clutter). Details about four SO-specific norms are shown in the lower half of Table 1.

\section{Methodology}
There are two parts of this study: identifying violations of {\itshape generic} and {\itshape specific} norms in SO. To study the pattern of generic norms, we collected the data from the SO heat detector bot~\cite{noauthor_search_nodate} that identifies violations based on regular expressions and various machine learning algorithms. We collected all comments that were flagged by the bot between 16 May 2016 and 31 January 2020. There were a total of 56382 comments. Since these comments are related to violations of generic norms (norm type 1) and did not contain violations of SO specific norms (norm type 2), we created a deleted comments dataset, by collecting comments that were posted on a particular day (2 December 2019), and subsequently checking which of those were deleted from that set in the next two months. We discuss these in detail in the following sub-sections.

\begin{table}[ht]
\caption{Norm categorization}\label{tab1}
\resizebox{\textwidth}{!}{%
\begin{tabular}{|p{1.8cm}|p{2cm}|p{4cm}|p{4cm}|p{1.5cm}|}
\hline

\bfseries{Norm violation type} & \bfseries{Label} & \bfseries{Description} & \bfseries{Examples} & \bfseries{Evidence} \\ \hline

\multirow{18}{4em}{Violation of generic norms}  & Personal harassment & Personally targeting, name calling, intimidating, defaming or trolling one user. & Can you please stop posting such unrelated rubbish.& ~\cite{noauthor_code_nodate}  \\\cline{4-4}
    &  &  & Shut up sir..... &  \\ \cline{2-5}
 
 & Racial & Treating a user unfavourably based on his/her skin tone, race or ethnicity. & Somebody is blocking you, are you Chinese? & ~\cite{noauthor_code_nodate}\\ \cline{4-4}
    &  &  & Who the hell are you to talk? An Arabian terrorist? &\\ \cline{2-5}

 & Swearing & Usage of any swear words or profanities to abuse one. & That means you have fu***d-up the routers.& ~\cite{noauthor_code_nodate} \\ \cline{4-4}
    &  &  & This community is moving to a**hole level.&  \\ \cline{2-5}
 
 & Other unwelcoming & Any other unfriendly comments or accusations which violates the inclusive policy of SO. & You copied my idea.& ~\cite{noauthor_code_nodate} \\ \cline{4-4}
    &  &  & This is spam. &\\ \hline

\multirow{12}{4em}{Violation of SO specific norms} & Gratitude& Expressing gratitude. & Thanks for your response. & ~\cite{noauthor_no_nodate} \\  \cline{4-4}
    &  &  & Thanks, works perfectly. & \\ \cline{2-5}
 
 & Apologizing & Expressing apology for mistakes. & Sorry I was offline.& ~\cite{noauthor_no_nodate} \\ \cline{4-4}
    &  &  & My humble apologies sir.& \\ \cline{2-5}

 & Welcoming & Welcoming a new member to SO or to the community. & Welcome to Stack Overflow, Hamza. & ~\cite{noauthor_no_nodate} \\ \cline{4-4}
    &  &  & Welcome to SO Kyle!& \\ \cline{2-5}
    
 & No longer needed & These are comments that were once useful but are not anymore. But the specific reason is unknown. & I found your post valid. & ~\cite{noauthor_no_nodate}\\ \cline{4-4}
    &  &  & That would be a nice idea. & \\ \hline

\end{tabular}}
\end{table}

\paragraph{\bfseries Identifying generic norms and their violations - }In SO, a comment flagged by the bot may violate the abuse norms of SO. A sample entry from the bot is {\itshape ``SCORE: 7 (Regex:(?i)/bullshit NaiveBayes:1.00 OpenNLP:0.75 Perspective:0.90)"}. Here the Regex attribute shows the nature of the violation (i.e. the use of the word bullshit). While the entry has other attributes such as the SCORE indicating the extent to which a comment might be a violation (i.e. 7 out of 10) and the output from other machine learning algorithms such as NaiveBayes, we consider only those comments that have regular expressions. Only these can be classified into one of the four groups, as the actual comments have been deleted by SO. We only have the regular expression that was matched. After the bot flags a comment, the moderators review the nature of the violation. The outcome of the review process may be any one of the following: (1) the moderators may deem the comment to be appropriate and permit the comment to be available in the site, (2) the comment may be deleted from the site but the reason is not disclosed, (3) the author may remove the comment voluntarily, after the moderator contacts the author because of its inappropriateness, (4) the comment may be deleted by the moderator because of its inappropriateness. The classification of comments based on the outcome from the moderation process is shown in Table 2. Outcomes A and M can be considered as punishments for norm violation.

\begin{table}
\caption{Table describing outcomes of moderator review }\label{tab2}
\begin{tabular}{|l|l|l|l|}

\hline
Number &  Type & Description\\
\hline
1 & E &  Contains matching keywords, but not deleted (i.e. they  \textbf{E}xist in SO). \\
\hline
2 &  U & Deleted ( \textbf{U}nknown reason).  \\
\hline
3 &  A & Deleted voluntarily by the  \textbf{A}uthor. \\
\hline
4 &  M & Deleted by the  \textbf{M}oderator. \\

\hline

\end{tabular}
\end{table}

The dataset contained 56382 comments of which 673 comments do not have the links to locate the comment. In the remaining 55709 comments we found that only 19872 have Regex values to consider for norm classification. After collecting the set of Regex keywords used in SO from the GitHub repository of the heat detector bot\footnote{\url{https://github.com/SOBotics/HeatDetector}}, we manually classified the Regex collection into four norm groups --- personal harassment, racial, swearing and other unwelcoming comments as listed in Table 1. The full list of Regexs that correspond to each group can be found online\footnote{\url{http://www.mediafire.com/file/o39ypolha0v0ik0/Regex_classification.pdf/file}}. Then a  Python program was used to identify the outcome of each violation by clicking on the link for each comment and extracting the nature of the outcome from the resulting page (e.g. moderator deleted the comment). This extracted data was then grouped based on different outcomes.

\paragraph{\bfseries Identifying SO specific norms and their violations - } To identify SO specific norm violations, we collected comments from one particular day - 2 December 2019, for two months. And we identified that 3221 comments were deleted within that period. We examined the reason for moderation of the comments. The reasons fall under two major categories. The first one is pleasantries. The reason for moderation is that, being a technical question-answering site, SO  discourages the formalities of expressing gratitude, apology and welcoming someone new to the site or community. The second measure is the scale of essentiality of the comment. The reason of moderation is that the moderators may find those comments no longer needed towards knowledge sharing goal of the site. 
 
\section{Results}
This section presents results of norm violation for the two categories of norms. The discussion of these results is presented in the subsequent section.

\paragraph{\bfseries Violation of generic norms - }

Figure 2 shows the occurrence of the top fifty Regexs in the dataset. Out of the top ten Regex keywords, both swearing and personal targeting keywords appear thrice each and unwelcoming keywords appear four times. This shows that even in SO, a technical discussion forum, people may violate the norms of politeness and may tend to use abusive words so frequently. Moreover, the presence of keywords representing other unwelcoming comments show that people may accuse others or others' comments as being a spam, sarcastic or rude, which has been traditionally been considered as the duty of moderators. Racial abuse related keywords are found in small numbers (i.e. 377 comments). Complete Regex occurrence details can be found online\footnote{\url{http://www.mediafire.com/file/mez6z6lcszz6ybi/Regex_occurrence.pdf/file}}.

\begin{figure}[ht]
\includegraphics[width=\textwidth,height=6cm,keepaspectratio]{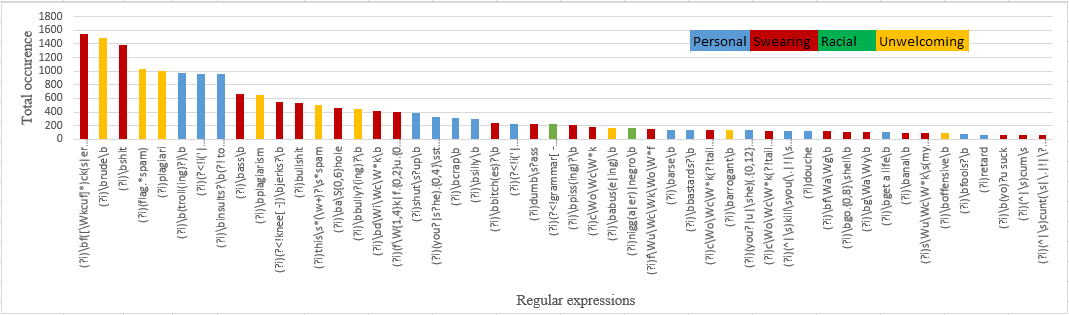}
\caption{ Top 50 Regexs and their corresponding occurrence} 
\end{figure}

Figure 3 shows the percentage of comments that violated norms belonging to the four categories. We have observed that personal harassment keywords and swearing terms are the two most common reasons for norm violation in the examined dataset (33.3\% and 33.2\% respectively). 66.5\% of the total violations come from both these two groups. However, another 31.5\% comes from other unwelcoming comments category. Only just 2\% of the violations were found for racial norms.

\begin{figure}[ht]
\begin{center}
\includegraphics[width=\textwidth,height=6cm,keepaspectratio]{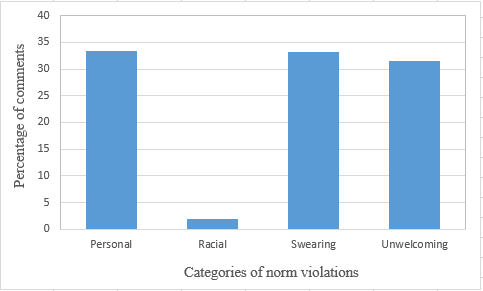}
\caption{Percentage of norm violations in four categories} 
\end{center}
\end{figure}

Figure 4 shows the outcome of the flagging process of the bot. As presented earlier, the outcome is one of the four options shown in Table 2.  We observed that in all four categories of norm violations, the type U outcome occurs the most, showing that the comments have been deleted from the site for an unknown reason. It is likely that the authors of the comments (without any prompting) realized the issue with their own posts and removed them. Out of 19872 comments evaluated, 45\% of comments belong to this category. The next category of outcome is those comments deleted by the moderator (M) with 28\%. Also, in 9\% of comments, the author voluntarily removed them (type A). The rest of the comments (18\%) are still present on the site (type E)  showing that despite possessing certain objectionable words, the human tolerance for these comments have not been exceeded. On the other hand, strict punishments are imposed on norm-violating comments. This is evident from the moderation process which removed 37\% of total comments which fell in categories A and M.

\begin{figure}[H]
\begin{center}
\includegraphics[width=\textwidth,height=6cm,keepaspectratio]{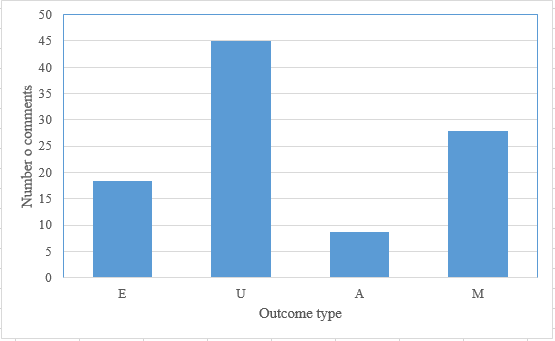}
\caption{Outcome statistics of bot flagging process} 
\end{center}
\end{figure}

Figure 5 shows the outcome of the flagging process by the bot for the four norm categories. It is evident that in all four categories, most of the comments are deleted from the site for an unknown reason (type U). Followed by this is type M where the moderator has deleted all these comments. Voluntary removal of the comment by the author (type A) is the outcome with the smallest count. It can be observed that the type E (comments still exist) outcome happens more than the type A outcome. Therefore, the general trend is U \textgreater  M  \textgreater   E  \textgreater  A in all four norm categories.

\begin{figure}[ht]
\begin{center}
\includegraphics[width=\textwidth,height=6cm,keepaspectratio]{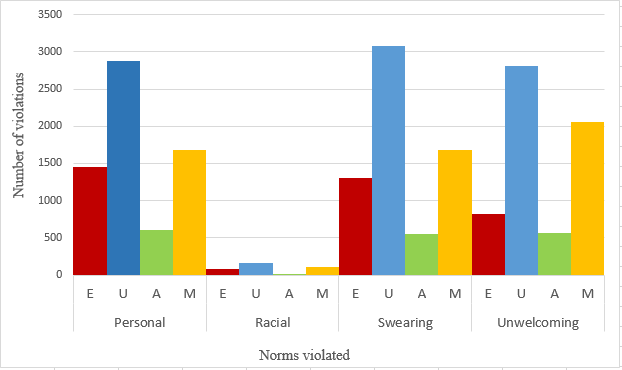}
\caption{Norm violations identified in bot flagging outcome} 
\end{center}
\end{figure}
\paragraph{\bfseries Violation of SO specific norms - } Figure 6 presents the percentage of various SO specific norm violations in the dataset. Out of 3221 comments evaluated, 84\% of the comments were in the `no longer needed' category. Since these are no longer needed for the site, those are deleted. 2\% are apologies that were removed. 2\% are welcome messages and 12\% are gratitude messages. It is interesting that pleasantries (apologies, welcome and gratitude) account for 16\% of the comments and these are in fact deemed to be not useful to the knowledge creation process. In addition to these, we also observed 9 comments in the dataset that were personal harassment messages indicating the violation of generic norms (not shown in Figure 6). This shows that the moderation process is not instantaneous (i.e., removing harassment messages takes time) since these are possibly milder (borderline) offenses and may require multiple users flagging them before a decision could be made.

\begin{figure}[ht]
\begin{center}
\includegraphics[width=\textwidth,height=5.5cm,keepaspectratio]{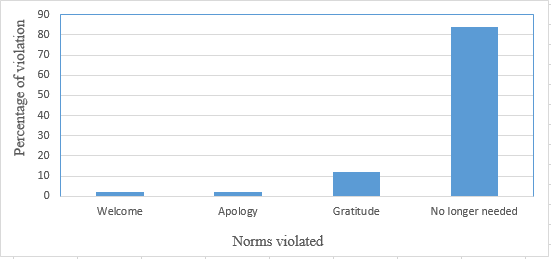}
\caption{Norm violations present in extracted dataset of one particular day} 
\end{center}
\end{figure}

\section{Discussion}
In this section we provide a discussion of the results presented in Section 5, particularly discussing their implications for developing a recommendation system to prompt the users to follow the norms of SO when posting a comment.
\paragraph{\bfseries Generic norms - } From the results shown in Section 5, it can be inferred that nearly equal contribution for generic norms violation come from the three categories of personal, swearing and other unwelcoming comments. Also, these three together contribute substantially towards total generic norm violation of 98\%. Therefore, if we can restrict the abusive language usage, we would be able to improve the quality of the system for everyone. This provides an opportunity to develop an online norm recommendation system for comments which may violate the generic norms of SO. Figure 7 displays the workflow of the application which provides norm recommendation.

\begin{figure}
\begin{center}
\includegraphics[width=\textwidth,height=8cm,keepaspectratio]{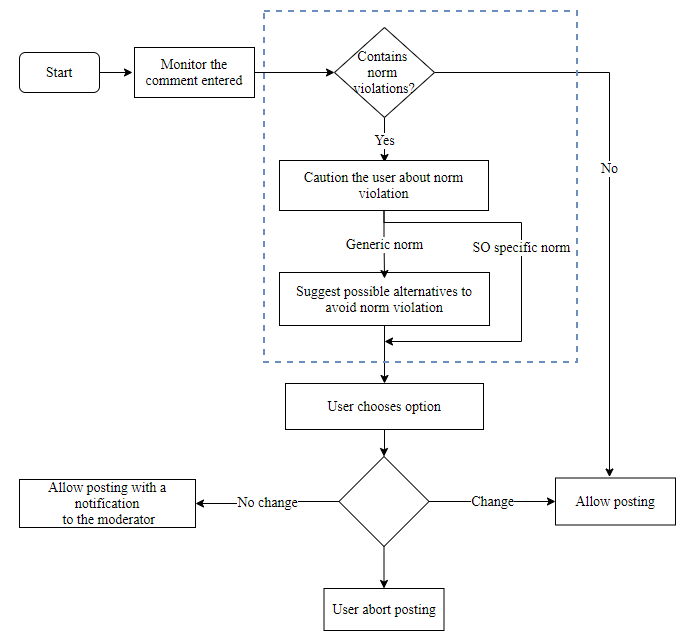}
\caption{The proposed norm adherence recommendation system } 
\end{center}
\end{figure}

In the recommendation system, when a new comment has been entered, the generic norm violations can be detected. If the comment violates norms, the system would alert the user regarding the abusive content and the nature of norm violation. Moreover, the system would provide certain rephrased options for the same comment using deep reinforcement learning techniques such as the ones that suggest code auto-complete~\cite{synced_deep_2019}. If the user accepts the proposal, the rephrased comment is posted. If the user does not rephrase or accept the options presented, the post would be allowed, however, a notification will be sent to the moderators regarding the norm violation. Then it would be the discretion of the moderator to review and decide the destiny of the comment. The user also has the option to abort the comment in which case the comment will not be posted.  Figure 8 shows an example of the options available to the user for rephrasing a bad comment.

\paragraph{\bfseries SO specific norms violations - } From the results shown in Section 5, it is evident that more than three-fourths of the comments in SO are in the category `no longer needed' which are abiding by the specific norms of SO and are neutral in nature. However, these comments are not contributing anything productive to the community, which is likely to be the reason for moderation. In addition to these set of comments, pleasantries also became a substantial reason for norm violations. Therefore, if we could limit the usage of these two categories of comments, SO specific norm adherence can be enhanced. Figure 7 shows that if a comment violates SO specific norms, the system would notify the user about the nature of norm violation. The user can abstain from posting the comment. If not, the user will be allowed to post the comment and the moderator would be notified regarding norm violation. 

\begin{figure}[ht]
\begin{center}
\includegraphics[width=\textwidth,height=6cm,keepaspectratio]{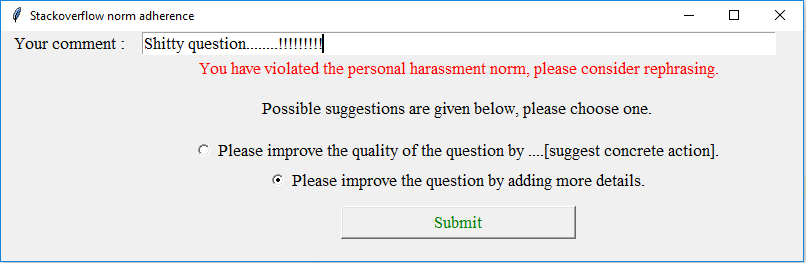}
\caption{Norm recommendation system } 
\end{center}
\end{figure}

In the future, we intend to build the norm recommendation system proposed in Figure 8 using deep learning techniques. Also, we plan to extend the proposed system to improve the reputation of users in SO community by abiding by the norms pertaining to the site. Thereby, better knowledge sharing without clutter can be facilitated, trust can be guaranteed and gentler treatment can be expected among community members.

\section{Conclusion}
The type of norms and their violations in SO are seldom addressed by prior work and that formed the focus of the current work. Our objectives are to identify and quantify the patterns of norm violations in SO comments and to propose a norm recommendation system for SO comments. We have identified two categories of norms, the generic norms and SO specific norms. We found that a significant proportion of violations in the first category has been contributed by the violation of three norms: personal harassment, swearing and other unwelcoming comments. In the second category, the main violations are `no longer needed' and pleasantries. We have proposed an approach that can identify and alert the user regarding the presence of violations in comments which would potentially limit norm violations in SO.

\printbibliography

\end{document}